\documentclass[journal]{IEEEtran}
\usepackage[lmargin=1.40cm, rmargin=1.40cm,tmargin=1.2500cm,bmargin=1.250cm]{geometry}
\usepackage{moreverb}
\usepackage{subfigure}
\usepackage{xfrac}
\usepackage{amsmath}
\usepackage{algorithm}
\usepackage[noend]{algpseudocode}
\usepackage{lineno,hyperref}
\usepackage{graphicx}
\usepackage{amsmath}
\usepackage{amsfonts}
\usepackage{graphicx}
\usepackage{bm}
\usepackage{amsbsy}
\usepackage{tikz}

\begin{document}
\title{Vulnerability and Efficiency Assessment of Complex Power Grids Using Current-Flow Line Centralities }
\author{Somnath Maity$^1$~\IEEEmembership{Senior Member,~IEEE,}\thanks{$^1$ The author is with the Department of Electrical Engineering, National Institute of Technology, Rourkela, India (Email: somnathm@nitrkl.ac.in, somnatheeiitkgp@ieee.org)} and Premananda Panigrahi}
\maketitle
\begin{abstract}
The centrality measure (CM) is one of the most fundamental metrics for evaluating the efficiency and vulnerability analysis of complex power grids (CPGs). Despite an abundance of different CMs for individual nodes, there are only a few metrics available in the literature to measure the centrality of individual lines. We propose here the current-flow (CF) line CMs to identify the ranking of lines, where each set of lines is associated with a different level of importance. We then find the CMs using effective resistance and apply it to identify the important lines in a commonly used IEEE 118-bus network. Finally, the efficiency and vulnerability of CPG are analyzed to validate the proposed concepts.
\end{abstract} 
\IEEEpeerreviewmaketitle
\section{Introduction} \label{introd}
Cascading failures in complex power grid (CPG) networks are inherently large-scale network processes that cannot be satisfactorily understood from a  small-scale or local disturbance perspective~\cite{Song_2016}. The functioning of CPGs is crucially dependent upon the interconnections between network links or nodes~\cite{Wei_2019,Fang_2018,Liu_2018}. These interconnections are such that when some links/nodes in a power grid (PG) fail, others connected through them may also fail and the entire grid may collapse. In order to understand this robustness and design the resilient CPG networks, it is therefore necessary to know whether a CPG can continue to operate at a stable and reliable operation after a fraction of its nodes/links have been removed either through natural failures  or malicious attacks.  

It has, however, been reported that vulnerability and network robustness can be successfully analyzed by using the framework of percolation theory~\cite{Ricard_2008} in which the percolation phase transition occurs at some critical value of occupation probability.  Above this critical threshold, a giant component --- defined as a cluster that contains a significant proportion of the entire nodes in the network~\cite{Motter_2002,Crucitti_2004} --- exists. Below the threshold, the giant component is absent and the network is  collapsed. Moreover, the robustness of a CPG is crucially dependent upon the structure of the underlying network and the nature of the attack: random and intentional attack~\cite{Ricard_2008}. It is shown that a random network is robust against intentional and random  attacks. Whereas a scale-free network is highly robust against random failures but fragile  under intentional attack~\cite{Motter_2002}. This analysis thus provides important information about the increasing robustness of complex networks. Once the most important links/nodes are found, one can increase the PG's robustness by protecting the key components such as implementing novel control techniques~\cite{Xue_2018} or policies to secure the most important lines/nodes~\cite{Mohammed_2020,Qi_2015,Evangelos_2013} or, directing energy sources to preserve important components of the CPGs~\cite{Mureddu_2016}. In this brief, we explore the application of robustness assessment tools~\cite{Qi_2015,Evangelos_2013} that can be deliberately used to guide the design of a CPG. 

However, to understand and design such a CPG, the identification of key transmission lines and their characterization are highly ­relevant to the network's structure and dynamics. There are several concepts and a growing number of metrics in network science --- such as centrality measures (CMs)~\cite{Boccaletti_2006} --- that can allow us to characterize the role of network nodes for structure and dynamics. There are, however, only a few measures for line centrality of CPG networks. Most of them are centered around the concept of betweenness centrality (BC)~\cite{Mureddu_2016}, other make use of the concept of bridging node degree line centrality~\cite{Wei_2019,Liu_2018} or are based on the spectrum of the network's Laplacian matrix using effective resistance distance (ERD)~\cite{Mieghem_2017}. We introduce here the concepts of current-flow (CF) closeness and betweenness centrality for lines and demonstrate that  line centrality provides additional information about the network constituents.  We also introduce the concepts of CF  spanning trees (ST) line BC for PG networks. This new metric --- corresponds to the fraction of minimum ST containing a particular line~\cite{Teixeira_2016} --- has a potential advantage not only to help in evaluating the network connectivity but also to evaluate how redundant a transmission line is in a given network.  For a given line, the ST centrality value can reflect whether the line removal can cause to disrupt the PGs or if there are some alternative ways to keep the grid connected, reflecting the redundancy of the network. 

We also apply the CF efficiency metric~\cite{Kai_Liu_2018} to calculate mean reciprocal of ERD between all pairs of nodes in the power network. This metric considers the multipath effect and is more suitable for measuring the efficiency and vulnerability assessment of real-world PGs. Moreover, it can identify the critical lines of the PGs even if the grid is disconnected. 
\section {Power Grid Network}
\subsection{PG as a Graph}\label{Sec:Graph}
To identify critical components using CF centralities, we first model the PGs as a weighted graph. From graph-theoretic point of view, the CPG with $n$ buses or nodes, and $L$ transmission lines or links can be represented by a weighted graph $\mathcal{G}=(\mathcal{N},\mathcal{L},\mathcal{W})$ along with the weight function defined by $\mathcal{W}\!:=\!\{w:\mathcal{L}\!\rightarrow \!\mathbb{R}^{+}$\}. In order to incorporate the electrical properties of PG networks, we further define a vector $\mathcal{X}\!:\!=\!\{jx_l:l\!\in\!\mathcal{L},j\!=\!\sqrt{-1}\}$, where  $jx_l$ is the reactance of a transmission line $l$ and also consider a weighted adjacency matrix $A$ to specify the interconnection pattern of the graph $\mathcal{G}$: $w_{ik}=1/|jx_l|$ only if the pair of nodes $i$ and $k$ are connected by a link; otherwise $w_{ik}=0$. This assumption is valid when transmission lines resistances $r_l$ are assumed to be negligible than the reactance values $|jx_l|$.

Under this formulation, the dc PF equation derived from nonlinear ac equation: $P_{i}\!=\!\sum_{k=1}^{n}  w_{ik} |V_{i}||V_{k}|\mbox {cos}(\delta_{i}-\delta_{k})$, can be approximately written as~\cite{Cetinay_2018}
\begin{eqnarray}
P_{i}&:=&\sum_{k=1}^{n} w_{ik}(\delta_{i}-\delta_{k})=\delta_i\sum_{k=1}^{n} w_{ik}-\sum_{k=1}^{n} w_{ik}\delta_k \label{node}
\end{eqnarray}
for $|V_{i}|=|V_{k}|=1~\mbox{p.u.} \;\mbox{and} \;(\delta_{i}-\delta_{k}) \to 0$. Here,  $P_i$ is the active power injected at bus $i$, $w_{ik}$ is the line admittance between the buses $i$ and $k$, and $V_{i}$, $V_{k}$, $\delta_i$, and $\delta_k$ are the corresponding bus voltages, and voltage phase angles at bus $i$ and $k$, respectively. Since the Eq.~(\ref{node}) holds for every bus $i$, the net active power injected into the network can be expressed as
\begin{equation}
 P=(D-A)\delta  \label{node_net}
\end{equation}
where $A=\sum_{k=1}^{n} w_{ik}$, $P$ = $[P_1 \;. . .\;P_n]^T$ is the vector of net active power injection at the nodes, $D=\mbox{diag}\sum_{k=1}^{n} w_{ik}$ is the weighted degree diagonal matrix, and $\delta$ = $[\delta_1\;. . .\;\delta_n]^T$ is the vector of voltage phase angle with weighted Laplacian matrix defined by $L=D-A$.

For a connected graph, $L$ is a symmetric positive semidefinite matrix~\cite{Nicolas_2020}. Thus, $L$ has a full set of $n$ real and orthogonal eigenvectors with real nonnegative eigenvalues $0=\lambda_1\leq\lambda_2\leq\lambda_3...1/\lambda_k...\leq\lambda_n$. Because of this zero eigenvalue, $L$ cannot be invertible and solution of (\ref{node_net}) cannot be obtained directly to find unknown  $\delta $ at each node for a given supply and demand.  
\subsection{Generalized Solution Using Spectral Decomposition}\label{Sec:Spectral}
However, using spectral decomposition~\cite{Mieghem_2017}, $L$ can be decomposed as $L=U\Lambda U^T$, where $\Lambda=\mbox{diag}(0,\lambda_2,\lambda_3...\lambda_n)$ and $UU^T\!=\!I\!=\!U^TU$. The columns of $U$ are eigenvectors of $L$ normalized in such a way that their length is one. In particular, the first column of $U$ is $u/\sqrt{n}$, associated with the all-one vector belongs to $\lambda_1=0$. Thus, the generalized inverse of Laplacian matrix $L$ can be expressed as: $L^{+}=U\mbox{diag}\left(0,1/\lambda_2,1/\lambda_3...1/\lambda_k...1/\lambda_n\right)U^T$. 
This matrix is symmetric and positive semidefinite. It has also zero eigenvalue associated with the all-one eigenvector $u$. The eigenvalues of $L^{+}$ are the reciprocal of the eigenvalues of $L$. Therefore, applying the projection of $L^{+}$ onto $L$, we obtain the product of two matrices: $LL^{+}=L^{+}L=I-\frac{J}{n}$, where $J=uu^T$ is the matrix of all ones. Now, substituting these products into $\left(L+\frac{J}{n}\right)\left(L^{+}+\frac{J}{n}\right)=I$ and satisfying the conditions: $Lu\!=\!0\!=\!L^{+}u$ and $u^TL\!=\!0\!=\!u^TL^{+}$, we can obtain the generalized or Moore-Penrose pseudo-inverse of $L$ as~\cite{Mieghem_2017} $L^{+}=\left(L+J/n\right)^{-1}-J/n$.

Here, the generalized inverse $L^{+}$ of $L$ is derived by inverting a suitable perturbed version of $L$ and then subtracting the perturbation $J/n$, representing the average phase angle of the bus voltages $\delta_{\rm avg}$. By choosing $\delta_{\rm avg}=0$ as the reference phase angle, it is easy to compute the network's reactance matrix $L^{+}$ --- also called as effective resistance distance (ERD) matrix~\cite{Mieghem_2017}; and thus solve the dc PF equations: $\delta=L^{+}P$. 
\subsection{Evaluation of ERD and Its Impacts}\label{Sec:RD}
The ERD between a pair of nodes in a resistive network is defined as the potential difference between those nodes when a unit current is injected at one node and leaving out from the other~\cite{Ghosh_2008}. While in power network, there are several load and generator buses and, under dc PF approximations, the active PFs over the transmission network resulting in the phase angle differences. This analogy enables the introduction of the concept of ERD  $Z_{ik}$ to capture the relation between the injected active power and voltage phase angles as $\delta_i-\delta_k\!=\!Z_{ik}P_{ik}$, where $P_{ik}$ is the active PF from node $i$ to $k$, and  $Z_{ik}\!=\!1/w_{ik}$ is the line reactance between these nodes. Now, combining this equation with $\delta=L^{+}P$, we obtain the resultant angle difference
\begin{eqnarray}
\delta_i-\delta_k:=\left(v_i-v_k\right)^T\delta=(v_i-v_k)^TL^{+} P_{ik}(v_i-v_k)
\label{Eq:Resis}
\end{eqnarray}
where $v_i$ and $v_k$ are the orthonormal column vectors of $U$ with $v_iv_k=1$ if $i=k$, else 0. This can be further simplified to obtain the ERD between nodes $i$ and $k$ as:
\begin{eqnarray}
Z_{ik}=(v_i-v_k)L^{+}(v_i-v_k)^T=L_{ii}^{+}+L_{kk}^{+}-2L_{ik}^{+}.
\label{Eq:ERD}
\end{eqnarray}

Here, ERD $Z_{ik}$ provides a precise interpretation of the elements of pseudo-inverse $L^{+}$. The diagonal element $L_{ii}^{+}$  or $L_{kk}^{+}$ indicates about the closeness of node $i$ or $k$ with respect to the rest of the network, respectively. While off-diagonal element $L_{ik}^{+}$ tells about the resistance proximity of the nodes $i$ and $k$. It is high if the nodes are close, e.g., their ERD is low; and low if they are in  distant apart, that is, their ERD is high. Moreover, such a RD proximity can be further summed-up for all possible pairs of nodes to obtain the total effective RD $Z_i$ and thereby the mean distance of the graph $d_g\!=\!Z_i/n$. By using (\ref{Eq:ERD}) and spectral decomposition of $L^{+}$, we can easily calculate the $Z_i$ with respect to node $i$ as~\cite{Mieghem_2017}:
\begin{eqnarray}
Z_i &=&\frac{1}{2}\sum_{i=1}^{n}\sum_{k=1}^{n}Z_{ik}=\frac{n}{2}\sum_{i=1}^{n}\sum_{k=1}^{n}L_{ii}^{+}+L_{kk}^{+}-2L_{ik}^{+} \nonumber\\
&=& \sum_{i=1}^{n}L_{ii}^{+}=n\mbox{Tr}\left(L^{+}\right)=n\sum_{i=2}^{n}\frac{1}{\lambda_i}
\label{Eq_Z_avg}
\end{eqnarray}
since $L_{ik}^{+}:=u^TL^{+}u=0$. Note that ERD (\ref{Eq_Z_avg}) between two nodes is proportional to the arithmetic mean of the eigenvalues of $L^{+}$, which is the reciprocal of the harmonic mean of the eigenvalues of the Laplacian matrix. The second smallest eigenvalue of $L$ is known as the algebraic connectivity of the network~\cite{Mieghem_2017}) and it is a measure of how easily the network can be divided into two disconnected components~\cite{Nicolas_2020}: the closer the eigenvalue $\lambda_2$ to 0, the easier the network to be divided. In particular, if $\lambda_2 \to 0$, then $1/\lambda_2$ is very large. Thus, the networks with larger mean ERD among the nodes  are easily separable, and it is typically determined by the magnitude of smaller eigenvalues of $L$~\cite{Zhan_2010}. 

In addition, the network analysis using ERD allow us to efficiently calculate the CMs and redistribution of powers when generators, loads, and topological structures of CPGs are changed under various natural failures and/or malicious attacks.
\section{Evaluation of CMs Using ERD} \label{Sec.CMs_RD}
\begin{figure}[t]
\centering
\includegraphics[width=3.250 in]{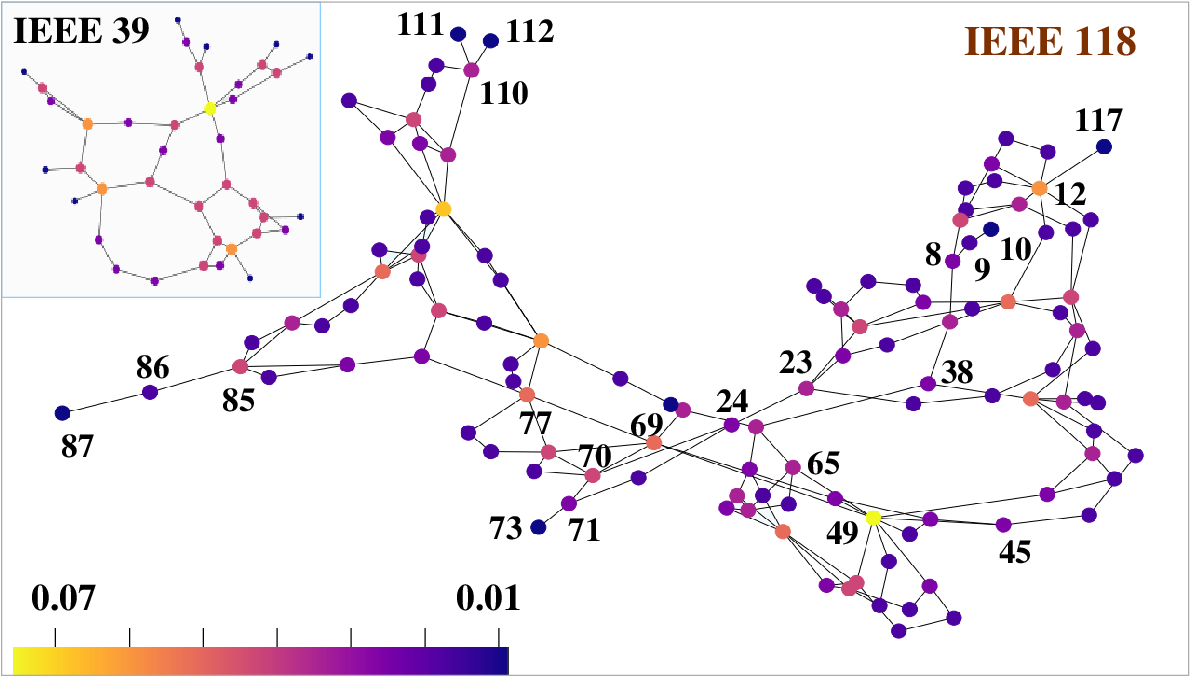}
\caption{Simulated network structures of wighted IEEE 118 and 39-bus systems. Color-map represents the node degree distribution of IEEE 118-bus network.} 
\label{Topo_IEEE}
\end{figure}

In this section, we examine the important line CMs and discuss how these centralities can be characterized by ERD to analyze the CPG network~\cite{IEE_Bus} (Fig.~\ref{Topo_IEEE}).
\subsection{CMs Using ERD} \label{Sec.CMs}
\subsubsection{Degree Centrality}
\begin{figure}[tbh]
\centering
\includegraphics[width=\columnwidth]{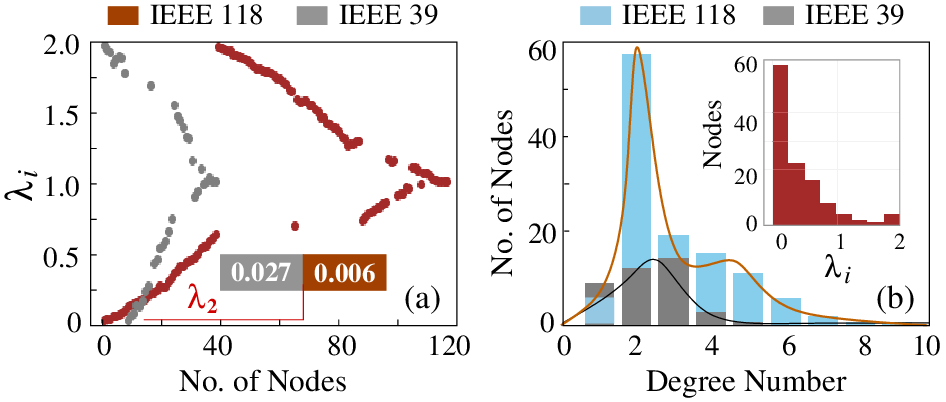}
\caption{Normalized Laplacian eigenvalues $\lambda_i$ and node degree distributions of two weighted PGs. (a) Spectral of normalized $\lambda_i$. (b) Similarity between node degree and normalized $\lambda_i$ distributions with  smallest nonzero eigenvalue $\lambda_2$.} 
\label{Degree}
\end{figure}

Degree centrality is the simplest CM to rank the nodes of a CPG according to its node degree. This centrality represents the connectivity of a node to the rest of the network and reflects the immediate effects of that node to exert its influences to the rest of the network. However, the connectivity of a node in PG network is not related to its geodesic distances; rather, it is related to the shortest electrical distance defined by line admittance or reactance value between the nodes. Therefore, by using the concept of average graph ERD (\ref{Eq_Z_avg}), the degree centrality of a node $i$ can be represented as $D_c=Z_i/n$ or,
\begin{equation} 
D_c:=\frac{1}{n-1}\sum_{i=1}^{n}L_{ii}^{+}=\frac{1}{n-1}\sum_{i=2}^{n}\frac{1}{\lambda_i};\;\mbox{since} \; L_{11}^{+}=0.
\label{Eq:D_c}
\end{equation}
where $n\!-\!1$ is the normalization factor.  Note that effective graph resistance characterizing $D_c$ can be determined here by the magnitude of small eigenvalues of $L$, which correspond to the large eigenvalues of $L^{+}$. In past, it has been shown that the largest eigenvalues of $L$ are mainly determined by the largest-degree node, while the smallest nonzero eigenvalue of $L$ depends on the way the nodes are connected. In fact, there is a strict similarity between the Laplacian eigenvalues  $\lambda_i$ and the degree distributions of nodes~\cite{Zhan_2010}. Here, we calculate the normalized  $\lambda_i$ and find the magnitude of small $\lambda_i$ to evaluate the eigenvalues distributions as shown in Fig.~\ref{Degree}. It shows that both these distributions are very similar and, as the number of nodes increases, the distributions become increasingly peaky and more accurate.
\subsubsection{CF Closeness Line Centrality}
Let us now bridge the lines and cut-off nodes to be eliminated to isolate the PGs into two or more disconnected components. Such elements can assess the grid vulnerability by irrupting  active PFs of the system due to its discontinuity. The disconnected component centrality of a line --- also termed as CF closeness line centrality~\cite{Brandes_2005}) --- can be easily measured by using the ERD. It is a measure of mean distance from a node to the other nodes. A central node with respect to $C_c$ is a node that is separated from others by only a shortest mean ERD. Now, defining the mean distance of node $i$ from the other nodes as $d_i=\sum_{k=1}^{n}Z_{ik}/n$ and substituting (\ref{Eq:ERD}) into it, we obtain the normalized $C^{(i)}_c$ for node $i$ as
\begin{equation}
C^{(i)}_c:=\frac{n-1}{d_i}= \frac{n-1}{\sum_{k=1}^{n}Z_{ik}}=\frac{n-1}{nL_{ii}^{+}+\mbox{Tr}\left(L^{+}\right)}
\label{Eq:C_c}
\end{equation}
where node-level component $L_{ii}^{+}$ indicates the closeness of node $i$; while proportion of component $\mbox{Tr}(L^{+})\!=\!\sum_{i=2}^{n-1}\frac{1}{\lambda_i}$ indicates the network-level interaction or a mean ERD~({\ref{Eq_Z_avg}). Therefore, for a node removal or fault, the larger the smallest connected component of $\lambda_i$, the larger the measure, indicating a more vulnerable node. In practice, such an indicator can also be used to identify the most vulnerable disturbance spreader links without network redundancy. In~\cite{Mieghem_2017}, authors proved that best spreading line between $(i,k)$ node pair can be found by minimizing the  ERD $Z_{ik}$ as: $\min_{i,k \in \mathcal{N}} Z_{ik}=\min_{i,k \in \mathcal{N}}\left(L_{ii}^{+}+L_{kk}^{+}-2L_{ik}^{+}\right)$. From this, one can find the closeness line centrality for a line $l\in(i,k)$ as~\cite{Timo_2022}
\begin{eqnarray}
 C^{l}_c\!=\!\frac{n\!-\!1}{\min_{i,k \in \mathcal{N}}Z_{ik}}\!\cong \!\frac{n-1}{\frac{1}{C^{(i)}_c}\!+\!\frac{1}{C^{(k)}_c}}=\frac{(n\!-\!1)C^{(i)}_cC^{(k)}_c}{C^{(i)}_c\!+\!C^{(k)}_c}.
 \label{Eq: BSN}
\end{eqnarray}
The Eq.~(\ref{Eq: BSN}) reveals that best spreader pair in which each node is individually optimally connected to all other nodes, but they are not  mutually strongly connected in the network. The ranking of best spreader links therefore can be attained by finding the minimum and maximum values of the diagonal and off-diagonal elements of $L^{+}$ respectively, or directly by using (\ref{Eq: BSN}).
\begin{figure*}[t]
\centering
\includegraphics[width=2.425 in]{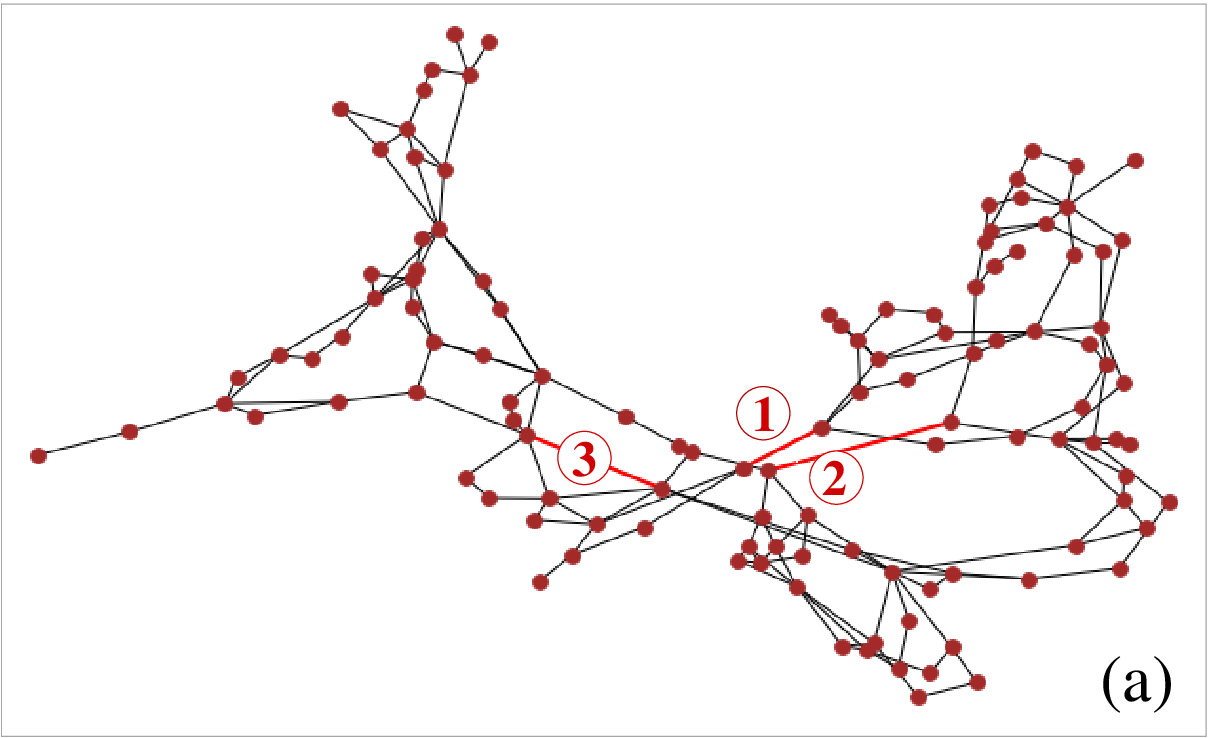}
\includegraphics[width=2.425 in]{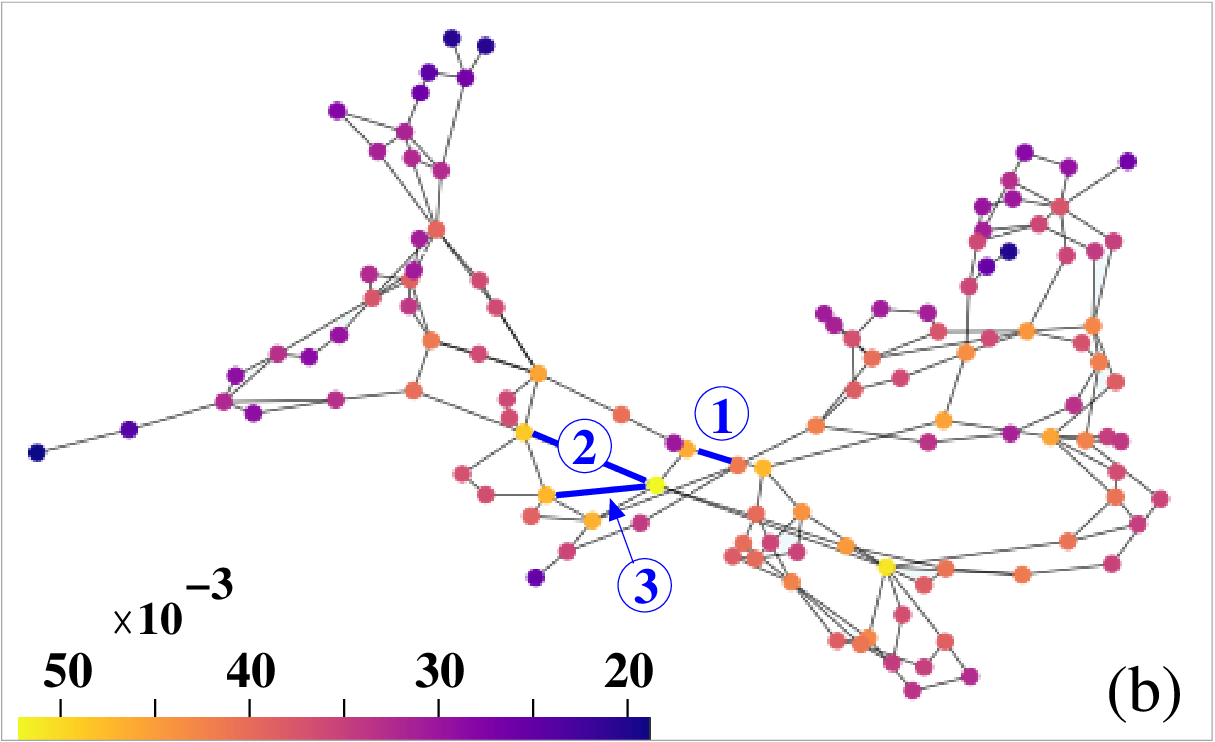}
\includegraphics[width=2.425 in]{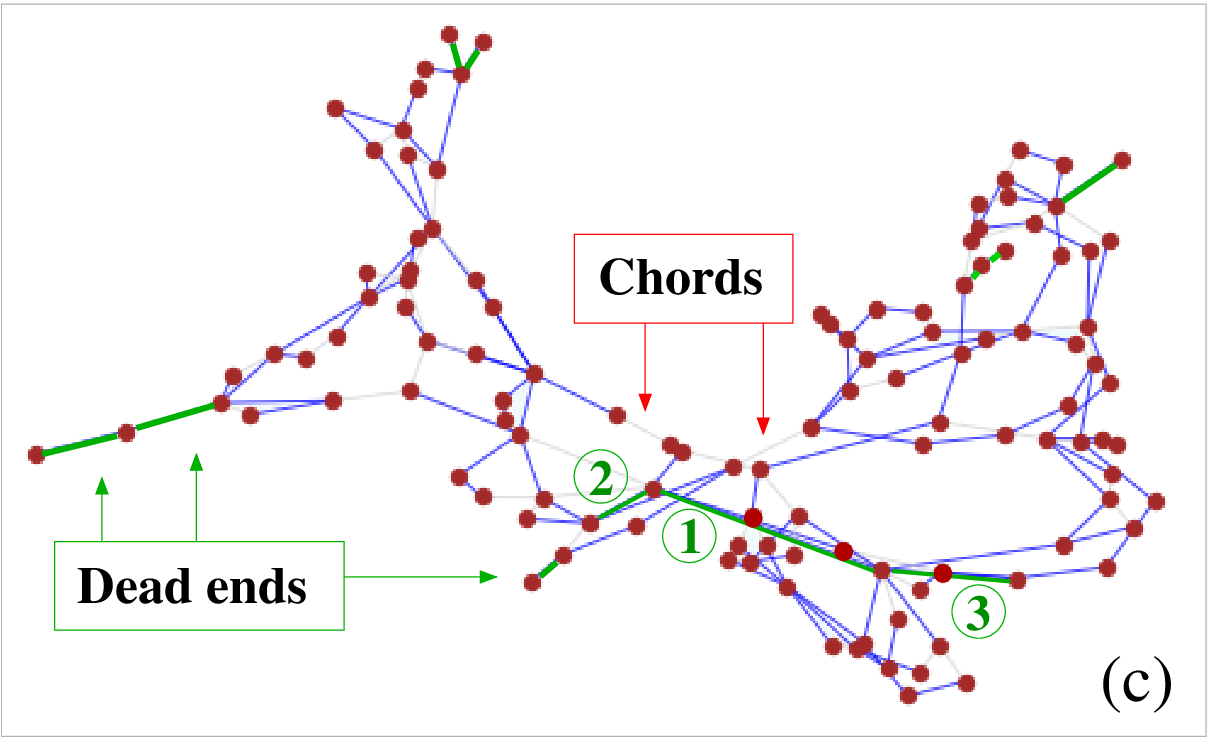}
\caption{Simulated network structures of IEEE 118-bus system showing the identification of highest central lines and their descending order ranking. a) Ranking of $B_c$ : $\small \textcircled{\bf \scriptsize{1}}\!:\!0.098\!\to\!\textcircled{\bf \scriptsize{2}}\!:\!0.091\!\to\!\textcircled{\bf \scriptsize{3}}\!:\!0.087$. (b) Ranking of $C^{(i)}_c$ (color map) and  $C^{l}_c$: $\small \textcircled{\bf \scriptsize{1}}\!:\!0.0106\!\to\!\textcircled{\bf \scriptsize{2}}\!:\!0.0104\!\to\!\textcircled{\bf \scriptsize{3}}\!:\!0.0102$. (c) Identification of MST and its ranking with dead ends links (green lines). The connecting nodes for all these important links are same as shown in Fig.~\ref{Topo_IEEE}.} 
\label{CMs_118}
\end{figure*}
\subsubsection{CF  Line BC}  
In addition, there exits an another important CM such as the BC which can quantify the power flow in the PG networks based on the shortest resistance paths between the nodes. 
To formulate this CF line BC, $B_c$, let us  consider the graph $\mathcal{G}$ with source node $s$  for the input power and a target node $t$ for the output power. The PF between this source-target pair $(s,t)$ passing though the transmission line $(i,k)$ is given by~\cite{Ghosh_2008}
\begin{eqnarray}
P_{ik}^{st}\!=\!\sum_{k=1}^{n}A_{ik}|\delta_i^{st}\!-\!\delta_k^{st}|\!=\!\sum_{k=1}^{n}A_{ik}|L_{is}^+- L_{it}^+-L_{ks}^{+}+L_{kt}^+|
\label{Eq:Flow_1}
\end{eqnarray}
where $\delta_i^{st}\!=\!L_{is}^+-L_{it}^+$ and $\delta_k^{st}\!=\!L_{ks}^+-L_{kt}^+$ are the phase angle of nodes $i$ and $k$ with corresponding pseudo-inverse of $L$.  This can be further simplified and averaged over all possible source-target node pairs $(s,t)$ to get the desired normalized $B_c$  as
\begin{eqnarray}
B_c=\frac{2\sum_{s<t}P_{ik}^{st}}{n(n-1)}.
\label{Eq:B_c}
\end{eqnarray}
Note that Eq.~(\ref{Eq:B_c}) is equivalent to the notion of random-walk BC for the large-scale network systems~\cite{Newman_2005}.

\subsubsection{Current-Flow ST Centrality}  
In summary, the ERD of a network is a graph metric that reflects PF capability of the network: the smaller the ERD, the better the network to transmit power. It is also related to the eigenvalues of the Laplacian matrix and ST of the graph~\cite{Mieghem_2017}. This measure is defined as the fraction of minimum ST (MST) in which a particular line is present, assessing the importance of that line in the network connectivity~\cite{Teixeira_2016}. For a weighted graph $\mathcal{G}$, the ST centrality of a line $l\!\in\!(i,k)$ can be mathematically computed as:
\begin{equation}
S_c=\frac{\tau_{l}}{\tau}; \;\; \mbox{for}\;\;\tau=\sum _{l\in \tau}\prod_{l\in\tau}w_{ik}\;\;\mbox{and}\;\;\tau_{l}=\sum _{l\in \tau_l}w_{ik}
\label{Eq:S_c}
\end{equation}
where $\tau\!\subset\!\mathcal{L}$ is the total number of different MST in a graph $\mathcal{G}$, and  $\tau_{l}$ is the number of different MST containing the line $l\in (i,k)$. This number $\tau$ of weighted ST can be expressed in terms of the Laplacian eigenvalues as $\tau=\frac{1}{n}\prod_{i=2}^{n} \lambda_i$~\cite{Mieghem_2017}; from which one can obtain MST so that  $\sum _{l\in \tau}w_{ik}$ is minimum among all the ST. While $\tau_{l}$ can be calculated as the total weight sum of that particular MST. As a result, $S_c$ can be used in confidence evaluation of the PGs  and it can allow to identify those lines that are more relevant for network connectivity, and how removing them leads to disruption in network structure.
\subsection{Numerical Analysis} \label{Sec.CMs_Ana}
In order to illustrate the importance of these CF centralities, we consider the IEEE~118-bus system as a benchmark network~\cite{Cetinay_2018} and evaluate the relative importance of its links according to the different CMs. The correlations of these CMs are also analyzed to identify the most significant lines. 

Fig.~\ref{CMs_118} shows that the most important vulnerable lines are different depending on the computed CMs and, each of these measures has some different aspect of PF distribution over the PG network. Compared to $C^{l}_c$, the line BC $B_c$ identifies the links that are more vital to the interconnection of geodesically distant apart nodes [Fig.~\ref{CMs_118}(a)]. Thus, it  facilitates the PF distribution between the distant sections of the PG networks. While closeness line centrality $C^{l}_c$ measures to rank the most disturbance spreader lines whose removal/attack could potentially cause to trigger the blackouts of the systems~\cite{Mieghem_2017})~[see Fig.~\ref{CMs_118}(b)]. However, despite the unique advantages, both of these measures do not allow us to infer any clear information about the network structure. They are all about giving information on how much active power can flow through a line in the shortest resistance paths, while $S_c$ is about the significance of a line, potentially identifying the lines that can break the network and reflecting if the network has a weak or strong redundancy. This measure finds the MST and identifies the lines whose removal can lead to an isolation of generator/consumer node from the grid including dead ends [Fig.~\ref{CMs_118}(c)]. 

Moreover, to further investigate the significance of $S_c$, we compute this  centrality and correlate its distribution with other centralities $C^{l}_c$ and $B_c$; and then, measure the line BC in MST. Fig.~\ref{CMs_Com} shows that $S_c$ has a different expression than the other CMs. It takes the value 1, expressing the direct importance of a line in a network; while all other CMs take their values between 0 and 1. This represents that there are some other lines that can keep the network connected, i.e., there is a different shortest resistance path for the CPG network, expressing the redundancy of a line. However, due to absence of any redundancy in MST, if a link is disconnected, the PFs in the grid is disrupted. This cannot be directly identified using the other CMs.
\begin{figure}[t]
\centering
\includegraphics[width=\columnwidth]{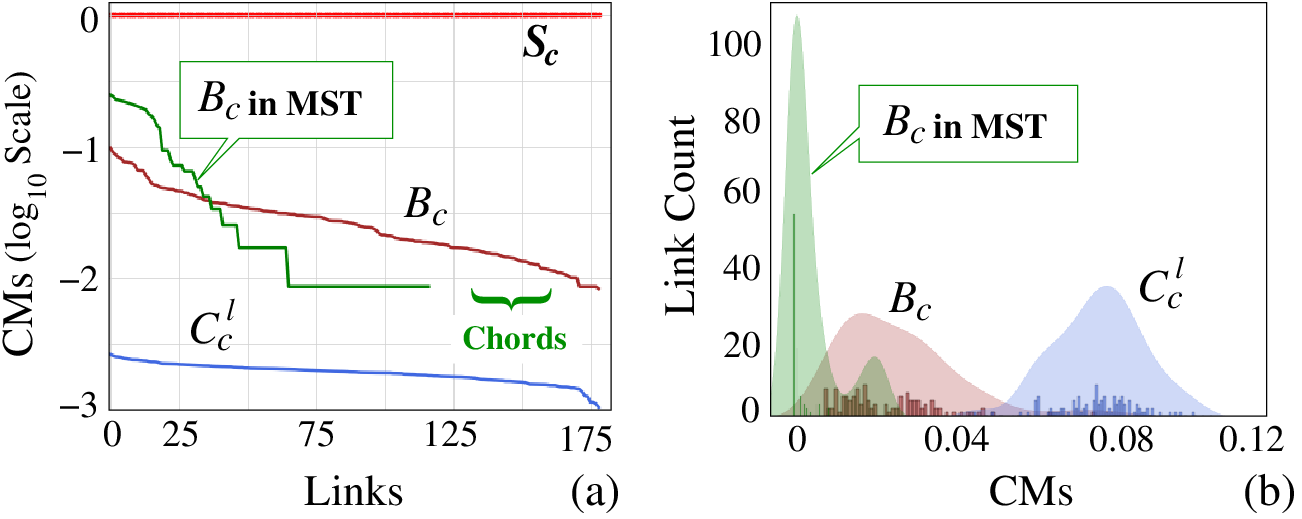}
\caption{CF line centralities and their distributions showing the significance in network structure. (a) Computed line CMs. (b) Distributions of links under descending order CMs. Centralities $B_c$ and $C^{l}_c$ with  small measured values infer no direct information about network structure and line redundancy. } 
\label{CMs_Com}
\end{figure}
\section{Vulnerability Analysis}
Network vulnerability is a property of complex networks that can quantify the susceptibility of information flow under the removal of a particular element in the graph. For a PG network, it is related to the drop in efficiency of the power transmission capability~\cite{Crucitti_2004}. The efficiency metric can be used in power networks to identify vulnerable components that can potentially trigger the systems to a cascading failures. Out of many definitions of network efficiency, we consider the global efficiency for weighted PGs which is mainly based on the inverse of shortest average ERD between any pair of nodes $(i,k)$, $E_g\!=\!\frac{1}{n(n-1)}\sum_{i\neq k}^{n}\frac{1}{Z_{ik}}$~\cite{Kai_Liu_2018}. We then use it to perform the vulnerability analysis of IEEE 118-bus network under the line attacks. The relationship between $E_g$ and vulnerability index $V_g(C)$ is established according to the CMs (Sec.~\ref{Sec.CMs}) as
\begin{eqnarray}
V_g (C)=\frac{E_g- E_g(C)}{E_g}\;\; \in[0 \;1]
\label{eq:vul}
\end{eqnarray}
where $E_g$ and $E_g(C)$ are the corresponding network efficiency computed before and after the removal of highly central lines. In order to evaluate these metrics, we successively remove the line according to the ranking of CMs, targeting to the most vulnerable line at a time. The strategy is numerically evaluated here by using the following greedy algorithm. Consider a weighted graph $\mathcal{G}\!=\!(\mathcal{N},\mathcal{L},\mathcal{W})$. At each iteration, (\ref{eq:vul}) is computed for all possible lines $(i,k)\in \mathcal{L}$ in the graph $\mathcal{G}$, and then select the line $l\!\in\!(i,k)$ that maximizes the vulnerability index,  $V^{l}_g(C)\!=\!\mbox{argmax}_{l \in \mathcal{L}}V_g(C)$. We then remove the selected line $l$ from the graph $\mathcal{G}^{\prime}=(\mathcal{N},\mathcal{L},\mathcal{W}\backslash l)$ and update the graph $\mathcal{G} \leftarrow \mathcal{G}^{\prime}$ for the  next iteration. The network efficiency $E_g(C)$ is recalculated at every iteration after the removal of a line. 
\begin{figure}[t]
\centering
\includegraphics[width=3.4 in]{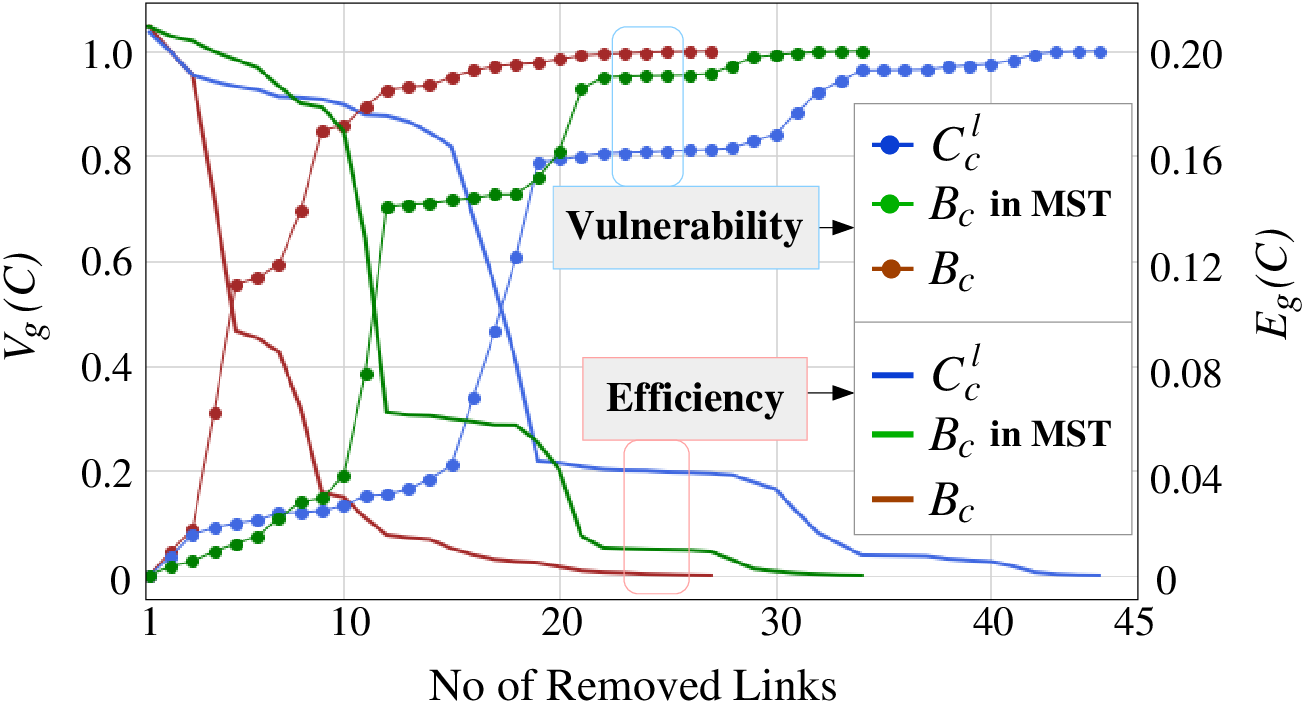}
\caption{Measured efficiency and vulnerability of IEEE 118-bus network with total number of nodes $n=118$ and lines $L=179$.} 
\label{Eff_Vul}
\end{figure}

The numerical results (Fig.~\ref{Eff_Vul}) reveal that network efficiency is considerably more sensitive to the attacks on 2-6\% most important lines with respect to the line BC, $B_c$. It is highly vulnerable to the attacks on the 8-12\% most important lines with respect to the CF closeness line centrality $C^{l}_c$, but less sensitive for the successive line removals. These results also reveal that PG network is highly vulnerable to attacks on important lines ranked with  $B_c$. However, for the few initial attacks (less than 5\%), the grid is less vulnerable to attacks on lines ranked with CF line BC in MST (i.e., $B_c$ in MST) compared to other the CMs $B_c$ and $C^{l}_c$, respectively. This is a consequence of bridging minimum ST path connecting all the network nodes including its dead ends. Therefore, removal of initial few central links are more likely disconnect the node [see Fig.~\ref{CMs_118}(c)] whose removal/isolation is less crucial to the overall power flow distribution and network efficiency. Consequently, after the initial attacks (e.g., greater than 5\% of the total lines), the network is highly disconnected and efficiency becomes more sensitive to the attacks on its central lines. The results are consistent with the analytical proof~\cite{Gago_2013} in which it has been shown that average line BC is higher in ST (or spanning subgraphs) and subsequently decreases in every stages of line removal. 

\section{Conclusion}
This study aimed to modify traditional centrality concepts for nodes to those for lines and assess the efficiency and vulnerability of CPGs. Instead of evaluating the geodesic shortest path between nodes, we used ERD to model the CF line centralities and found that the vulnerability to rank-based line attacks is related to the importance of a line in the PF distribution of the complex power network. We found that the CPGs are highly vulnerable to attacks on important lines according to BC. Meanwhile, ST centrality allows us to evaluate the confidence of the PGs and identify the lines that are more relevant for network connectivity. Removing these lines causes disruption in the network structure. The analysis using dependency between network efficiency and its vulnerability to line attacks thus gives a promising result for future studies on PG design and expansion. It can also be extended in order to design the grid to optimize the system’s PF and maximize its resilience/redundancy simultaneously.

\end{document}